\documentclass[%
  aps,
  preprint,
  amsmath,
  amssymb
]{revtex4-1}

\usepackage[T1]{fontenc}
\usepackage[utf8]{inputenc}
\usepackage{bm}
\usepackage{graphicx}
\usepackage[colorlinks=true,linkcolor=blue,citecolor=magenta,urlcolor=blue]{hyperref}

\usepackage{balance}

\usepackage[T1]{fontenc}
\usepackage[utf8]{inputenc}
\usepackage{bm}
\usepackage{graphicx}
\usepackage{slashed}

\hypersetup{
  colorlinks=true,
  linkcolor=blue,
  citecolor=magenta,
  urlcolor=blue
}
\usepackage{tcolorbox}
\tcbuselibrary{breakable}

\DeclareMathOperator{\Tr}{Tr}

\newtcolorbox{berryresult}{
colback=red!5,
colframe=red!70!black,
boxrule=0.8pt,
arc=2pt,
left=6pt,
right=6pt,
top=6pt,
bottom=6pt,
breakable
}

\begin{document}

\title{Asymptotic Quantum Gravity as an Infrared Geometric Theory
}

\author{J.~Gamboa}
\affiliation{Departamento de Física, Universidad de Santiago de Chile, Santiago, Chile}
\email{jorge.gamboa@usach.cl}

\author{N.~Tapia-Arellano}
\affiliation{Department of Physics and Astronomy, Agnes Scott College, Decatur, GA. 30030, USA}
\email{narellano@agnesscott.edu}


\begin{abstract}
We formulate the infrared sector of asymptotically flat quantum gravity
in terms of asymptotic configurations accessible to external observers.
Starting from the Regge--Teitelboim Hamiltonian that generates physical
evolution in the presence of gravitational constraints, we perform a
Born--Oppenheimer reduction separating slow asymptotic data from fast
bulk gravitational fluctuations. We show that integrating out the fast
sector induces a functional Berry connection over the space of
asymptotic charges, so that the effective infrared dynamics is governed
by parallel transport on this charge space.

In this framework, infrared gravitational states are naturally
organized into superselection sectors labelled by the holonomy of the
induced connection, and the reduced density matrix obtained after
tracing over ultraviolet bulk modes acquires a geometric contribution.
This provides an effective geometric description of the asymptotic
quantum gravitational sector, where quantization arises as a global
consistency condition under adiabatic transport rather than as a
spectral property of local bulk operators.
\end{abstract}


\maketitle

\section{Introduction}
The conventional formulation of quantum mechanics and quantum field theory is based on a fundamentally local description of spacetime \cite{Glimm:1987ylb,Haag:1963dh, Buchholz:1981fj}.  
Within this framework, the observer can be
idealized as external to the physical system, and time is defined locally in each
region of spacetime. Under these conditions, the quantum description is consistent,
predictive, and conceptually well established. However, once gravitation is taken
into account, the idealization of an external observer and a local notion of time is
no longer innocuous: gravity does not merely act as an additional dynamical field,
but instead determines the global causal structure of spacetime \cite{Dirac1958,ArnowittDeserMisner1962}.
 As a consequence,
the fundamental notions of observability, accessibility, and quantum state acquire
an inevitably global character.

From this perspective, it is natural to question whether genuinely quantum
gravitational effects should be sought primarily in local operators defined in the
bulk, or whether their physically relevant manifestation is instead encoded in
asymptotic observables. In general relativity, quantities such as energy, momentum 
and angular momentum admit an unambiguous definition only at infinity \cite{ArnowittDeserMisner1962,ReggeTeitelboim1974}
 and the
presence of horizons introduces causal restrictions that force the system to be
described in terms of reduced quantum states \cite{Hawking1975,Unruh1976}.
 These features suggest that a
quantum description of gravity must be formulated, at least in part, in terms of
asymptotic degrees of freedom.

This viewpoint naturally brings asymptotic gravity closer to the infrared sector of
gauge theories. In Yang--Mills theories, and in particular in quantum
electrodynamics, it is well known that physical states are not Fock states \cite{Chung1965,Kibble1968,KulishFaddeev1970}, 
 but
rather states dressed by clouds of soft modes \cite{Gamboa:2025dry,Gamboa:2025qjr,Gamboa:2025fcn}. These infrared degrees of freedom
are governed by gauge invariance and manifest themselves through geometric phases, holonomies \cite{Strominger2018}, and entanglement structures that persist even when local excitations
decouple. In this sense, the gravitational construction developed below should not
be viewed as ad hoc, but as the natural gravitational analogue of a mechanism that
is already well understood and under theoretical control in gauge theories.

It is important to emphasize, however, that asymptotic symmetries\cite{Bondi1962,Sachs1962,Strominger2014}
need not be taken
as the most fundamental starting point for describing the infrared sector. In gauge
theories, the same physical content can often be captured more directly in terms of
already dressed physical states, obtained through an adiabatic separation between
fast (ultraviolet) and slow (infrared) degrees of freedom. Within this approach,
asymptotic charges and symmetries emerge as effective descriptors of the same
underlying physics, rather than being postulated \emph{a priori} as organizing
principles of the state space.

In this paper, we propose that Berry phases and the adiabatic approximation\cite{Berry1984,ShapereWilczek1989}
provide a
natural and controlled framework to describe these infrared degrees of freedom, both
in gauge theories and in asymptotic gravity. Rather than starting from a complete
quantization of the bulk or from the Wheeler--DeWitt equation, we adopt a geometric
description of the space of configurations, in which physical states are obtained
through adiabatic transport and are characterized by functional holonomies. In this
framework, the relevant quantum information is not localized in pointlike
excitations, but encoded in the global structure of the dressed state and in its
infrared entanglement.

The advantage of this formulation is twofold. On the one hand, it allows for a
unified geometric treatment of well--known infrared phenomena---such as dressing,
memory effects, and the cancellation of divergences \cite{Strominger2017}.
 On the other hand, it suggests
an alternative interpretation of quantum gravitational effects, in which
ultraviolet physics is not directly observable but instead manifests itself
through infrared structures accessible to the asymptotic observer. In this sense,
quantum gravity emerges not as a local theory of the bulk, but as an effective
theory of the asymptotic sector, organized by symmetries, holonomies, and infrared
entropy.

From the present viewpoint, the quantum gravitational sector accessible to an
external observer should be regarded as an infrared effective theory defined
on the space of asymptotic configurations, in which the generator of physical
evolution is provided by Regge--Teitelboim boundary charges rather than by
local bulk operators.

 Our goal is not to
provide a complete quantization of the gravitational field, but rather to clarify
the physical role of asymptotic degrees of freedom and infrared structures in a
quantum theory of gravity, and to argue that genuinely quantum gravitational
features may be encoded in the global, infrared organization of physical states.

\section{Measurement in Quantum Gravity as an Infrared Theory}

The measurement process has occupied a central role in quantum mechanics since its
inception and is intrinsically tied to the notion of interaction between a quantum
system and a measuring apparatus. In standard formulations of the theory, this
interaction is described on the basis of a unitary evolution defined with respect
to an external time \cite{Kuchar1992,Isham1993}
 which allows the observer to be treated as an operationally
external agent relative to the observed system. However, this conceptual structure
relies crucially on the existence of an external notion of time and of a
well--defined evolution operator. When these notions are no longer available, as is
the case in theories where the geometry of spacetime is dynamical, the standard
formulation of the measurement process must be reconsidered. In such contexts, the
interpretation of measurement tends to shift from an internal dynamical process
toward a more relational description, whose physical meaning becomes clear only at
the asymptotic level.

This tension becomes particularly acute in the context of quantum gravity. In this case, spacetime dynamics cannot be regarded as a fixed background but instead form part of the quantum system itself, leading to the absence of a privileged notion of time and, consequently, of a global evolution operator. The quantum description then takes the form of a constraint condition---such as the Wheeler--DeWitt equation \cite{DeWitt1967}---rather than that of an evolution equation. Within this framework, the physical meaning of observables must be sought in relational terms and, naturally, in structures defined in the asymptotic regime \cite{Rovelli1991}.

This raises the question of how a measurement process can be defined in quantum gravity when the observer is inseparably tied to the bulk and cannot be operationally externalized. One possible approach is to shift the notion of the observer toward the asymptotic regime, as is done in asymptotic gravity. In this framework, the observer is effectively located at the boundary of spacetime and, although lacking access to local bulk observables, can nevertheless extract
physical information through infrared quantities and asymptotic structures that encode the gravitational dynamics in an indirect but physically meaningful way.

A concrete and conceptually clean illustration of these ideas is provided by quantum electrodynamics in the infrared regime \cite{KulishFaddeev1970,Chung1965,Kibble1968}. 
There, the role of asymptotic observers, infrared dressing, and the emergence of physical observables can be analyzed in a controlled setting, offering a useful guide for the gravitational case.

Following the standard logic of the Born--Oppenheimer (adiabatic) approximation, we separate the gauge field into slow and fast components associated with the infrared and ultraviolet sectors of the theory. Concretely, we write
\begin{equation}
A_\mu = A_\mu^{L} + A_\mu^{H},
\label{rep1}
\end{equation}
where the superscripts $L$ and $H$ denote light (slow) and heavy (fast) degrees of
freedom \cite{Weinberg1965}
 respectively. This decomposition is understood in the infrared sense,
with $A_\mu^{L}$ encoding the soft gauge modes that dominate the long--distance
dynamics, while $A_\mu^{H}$ collects the hard photon modes that admit a standard
Fock--space description.

The physical motivation behind this split is that the heavy sector adjusts
instantaneously to a given configuration of the soft fields, while the latter
evolve adiabatically in configuration space. This separation allows one to
systematically integrate out the fast degrees of freedom and to derive an
effective description for the infrared sector.

Adopting an adiabatic ansatz for the wave functional,
\begin{equation}
\Psi[A,\psi]
=
e^{\,i\,\gamma_C[A^{L}]}\,
\psi[A^{H},\psi],
\label{eq:adiabatic_ansatz_QED}
\end{equation}
the phase $\gamma_C[A^{L}]$ represents the Berry phase accumulated along an
adiabatic path $C$ in the space of soft gauge configurations. This geometric phase
encodes the infrared dressing associated with the slow modes.

Within this framework, the dynamics of the heavy sector is governed by an effective
Hamiltonian of the form
\begin{equation}
\hat H_{\mathrm{eff}}
= \int d^3 x \left[
\frac{1}{2}
\left(
-\,i\,\frac{\delta}{\delta A^{H}}
-
\mathcal A
\right)^2
+ V \right],
\label{eq:Heff_QED}
\end{equation}
where $\mathcal A$ denotes a $U(1)$ Berry connection induced by adiabatic transport
in the space of gauge configurations, and $V=V[A^L,A^H,\psi]$ is an effective
potential whose explicit form will not be required in the following.

\subsection{Evolution operator and infrared dressing}

At this stage it is useful to make an important conceptual digression. In the
Schr\"odinger representation, as in ordinary quantum mechanics, the time evolution
of the quantum state can be written formally as
\begin{equation}
\Psi[t]
=
\hat U(t,t')\,\Psi[t'] ,
\label{def1}
\end{equation}
where $\Psi[t]$ is a shorthand notation for the wave functional $\Psi[A,\psi;t]$.
The time--evolution operator is given by
\begin{equation}
\hat U(t,t')
=
T\,\exp\!\left[
-\,i\!\int_{t'}^{t} d\tau\, \hat H(\tau)
\right],
\label{eq:U_timeordered}
\end{equation}
with $T$ denoting time ordering. In the present case, where the Hamiltonian $\hat H$
is time independent, this expression reduces to
\begin{equation}
\hat U(t,t')
=
\exp\!\left[-\,i\,\hat H\,(t-t')\right].
\label{eq:U_timeindep}
\end{equation}

The definition (\ref{def1}) is appropriate for introducing asymptotic states in
situations where the evolution operator $\hat U(+\infty,-\infty)$ exists as a
well--defined object. In that case, $\hat U(+\infty,-\infty)$ coincides with the
$S$--matrix, and the asymptotic wave functionals $\Psi[\pm\infty]$ can be identified
with states in a Fock space.

In the infrared regime, however, this picture breaks down. The asymptotic states
are no longer Fock states, and the evolution operator must be supplemented by an
infrared dressing. In this case, $\hat U(+\infty,-\infty)$ naturally combines with
the Chung--Kibble--Kulish--Faddeev (CKKF) dressing \cite{Chung1965,Kibble1968,KulishFaddeev1970}
 yielding infrared--finite
asymptotic states.

A convenient and physically transparent way to implement this infrared dressing is
to exploit the universal separation between hard and soft photon modes that
characterizes infrared QED. This separation is precisely the one introduced in
Eq.~(\ref{rep1}), where the light component $A_\mu^{L}$ corresponds to the soft
photon sector, while $A_\mu^{H}$ describes the hard modes.

The soft sector contains modes with arbitrarily low energies that cannot be
described within a Fock--space framework. These soft modes couple universally to
the asymptotic charged currents, independently of the details of the hard process.
At leading infrared order, the interaction between soft photons and asymptotic
charges is governed by an eikonal Hamiltonian \cite{Weinberg1965}. As a consequence, the full
time--evolution operator factorizes into a hard contribution \cite{Yennie1961}
and a universal soft
factor, which can be evaluated exactly.

Integrating out the soft photon modes then amounts to attaching to each hard state
a coherent cloud of infrared photons. This construction yields the standard CKKF
dressing and defines the appropriate asymptotic states of the theory.

More explicitly, the time--evolution operator in the infrared may be written in the
factorized form
\begin{equation}
\hat U(t,t')
=
\hat U_{\rm hard}(t,t')\,
\hat W(t)\,
\hat W^\dagger(t'),
\label{eq:U_factorized}
\end{equation}
where $\hat U_{\rm hard}$ acts on the hard sector, while $\hat W$ is a Weyl
(displacement) operator that generates a coherent infrared photon state. For a set
of charges $e_a$ with asymptotic four--momenta $p_a^\mu$, this operator takes the
form
\begin{equation}
\hat W
=
\exp\!\left[
\int\!\frac{d^3\mathbf{k}}{(2\pi)^3\sqrt{2\omega_{\mathbf{k}}}}\,
\Big(
\alpha^\mu(\mathbf{k})\,\hat a^\dagger_\mu(\mathbf{k})
-
\alpha^{\mu*}(\mathbf{k})\,\hat a_\mu(\mathbf{k})
\Big)
\right],
\label{eq:CKKF_dressing}
\end{equation}
with the coherent amplitude fixed by the universal eikonal coupling,
\begin{equation}
\alpha^\mu(\mathbf{k})
\;\propto\;
\sum_a e_a\,\frac{p_a^\mu}{p_a\!\cdot\! k},
\qquad
k^\mu=(\omega_{\mathbf{k}},\mathbf{k}),\quad
\omega_{\mathbf{k}}=|\mathbf{k}|.
\end{equation}

Although the infrared dressing is usually introduced in operator language, it
admits a natural geometric interpretation. The family of coherent infrared vacua
generated by $\hat W$ forms a bundle over the space of asymptotic charge
configurations \cite{Strominger2017}. 
 Transport between the remote past and the far future induces a
pure geometric phase associated with the overlap of these vacua, which may be
written as the holonomy
\begin{equation}
\mathcal U_C
=
\mathcal P\exp\!\left(i\int_C \mathcal A\right),
\label{vestido}
\end{equation}
where $\mathcal A$ is a Berry connection on the space of infrared vacua. In this
sense, the CKKF dressing is equivalently described as a Berry holonomy generated by
integrating out the soft photon modes.

The relevant point in this interpretation is not the statistics of the soft photon
cloud by itself, which remains bosonic, but a global property of the physical
dressed state, obtained by combining the charged excitation with its infrared
cloud. The dressed state preserves the fermionic character of the underlying
charge, and this global feature constrains the allowed infrared holonomies. In
particular, consistency under parallel transport along a closed path $C$ selects a
non--trivial holonomy,
\begin{equation}
\mathcal U_C = -1 ,
\end{equation}
which implies the quantization of the associated Berry flux,
\begin{equation}
\oint_C \mathcal A = (2n+1)\,\pi .
\label{flux}
\end{equation}

As a consequence, the infrared--dressed Hilbert space decomposes into topological
superselection sectors characterized by this discrete invariant \cite{Buchholz1986}. States belonging
to different sectors cannot be continuously connected by local or infrared--regular
operations. In this precise sense the infrared--dressed states are topologically
protected, with the protection arising from the global structure of the infrared
configuration space and the Gauss law constraint, rather than from dynamical energy
gaps.

The same idea developed above can also be derived directly in the interaction
picture at the operatorial level. Consider the Dirac field in an external gauge
background,
\begin{equation}
i\,\partial_t \hat\Psi(t,\mathbf x)=\hat H_D[A](t)\,\hat\Psi(t,\mathbf x),
\qquad
\hat H_D[A]=\boldsymbol{\alpha}\!\cdot\!\big(-i\nabla-e\,\mathbf A\big)+\beta m+e\,\Phi,
\end{equation}
where $\hat\Psi$ is the Dirac field operator. Separating the electromagnetic field
into soft and hard components,
\begin{equation}
A_\mu = A_\mu^{\rm soft}+A_\mu^{\rm hard},
\end{equation}
one may treat the soft sector as adiabatically varying, while the hard sector
defines the fast degrees of freedom. Mutatis mutandis, the Born--Oppenheimer/Berry
ansatz implies that the effective evolution in the hard sector acquires a geometric
connection in the space of soft configurations. Equivalently, in the interaction
picture the effective infrared coupling can be written as
\begin{equation}
\hat H_{\rm int}^{\rm IR}(t)
=
\int d^3\mathbf x\;\hat J_\mu(t,\mathbf x)\,\mathcal A^\mu\!\big[A^{\rm soft}\big](t,\mathbf x),
\label{eq:Hint_Berry}
\end{equation}
where $\hat J^\mu=\bar{\hat\Psi}\gamma^\mu\hat\Psi$ is the conserved matter current
and $\mathcal A^\mu[A^{\rm soft}]$ is the Berry connection induced by integrating
out the fast sector. As a result, the infrared dressing is generated by the
corresponding holonomy,
\begin{equation}
\mathcal U_C
=
\mathcal P\exp\!\left(i\oint_C \mathcal A\right),
\end{equation}
so that the dressed state is obtained by parallel transport along the adiabatic
path $C$ in the space of soft gauge configurations.

\subsection{Charge quantization}

Asymptotically, the electric charge in QED is defined as a surface integral at
spatial infinity and is conserved as a consequence of Gauss’s law,
\begin{equation}
Q \;=\; \int_{S_\infty} dS_i\, E^i .
\end{equation}
This definition relies only on the asymptotic behaviour of the gauge field and is
independent of the detailed bulk dynamics. In the standard formulation, however,
Gauss’s law by itself does not account for the observed quantization of the
charge, which is typically associated with the underlying representation of the
gauge symmetry. It does not explain why the charge takes quantized values,
which are usually imposed as an external input.

In the present framework, the same asymptotic quantity admits a complementary
and intrinsically geometric interpretation. As discussed above, the infrared dressing 
endows the space of infrared configurations at $S_\infty$ with a Berry connection
$\mathcal A$, and physical dressed states are characterized by the associated
holonomy $\mathcal U_C$ defined in Eq.~(\ref{vestido}). The global consistency of
the dressed quantum state under closed transport in the infrared configuration
space then constrains the allowed values of this holonomy.

As indicated in Eq.~(\ref{flux}), this constraint leads to a quantization of the
Berry flux, which becomes a discrete topological invariant labelling inequivalent infrared 
sectors of the dressed Hilbert space. Since the Berry connection is directly tied to the asymptotic
electromagnetic field, this structure may be viewed as providing a geometric
origin for charge quantization: the same condition translates into a quantization of the admissible values of the 
asymptotic charge itself.

From this perspective, charge quantization emerges as a global
consistency condition on the infrared--dressed Hilbert space, rather than solely
as an independent input postulate.
While the asymptotic Gauss law guarantees conservation of the charge defined at 
$S_\infty$, the topology of the infrared configuration
space and the associated Berry holonomy offer a complementary way of
characterizing its allowed values, without invoking Dirac monopoles or 
singular bulk configurations.

\section{Infrared Quantum Gravity}

The purpose of the preceding discussion has been to establish, in a controlled
and familiar setting, how infrared dressing, adiabatic separation, and Berry
holonomies naturally arise in gauge theories. We now argue that the same
conceptual structure reappears, in a more subtle but equally robust form, in the
infrared sector of gravity.

We adopt a canonical description in which the notion of time is fixed by a
choice of foliation, as appropriate for an external (asymptotic) observer.
Concretely, we choose a proper--time parametrization of the slices by imposing
\begin{equation}
\dot N_\perp = 0, \qquad N_i = 0,
\end{equation}
so that the lapse is constant along the foliation and the shift vanishes
\cite{ArnowittDeserMisner1962,DeWitt1967}.
In this gauge the lapse plays the role of a (spatially dependent) Lagrange
multiplier for the Hamiltonian constraint, in direct analogy with the
proper--time gauge of the relativistic particle. This choice removes the local
arbitrariness in the definition of the evolution parameter while preserving its
physical role, and makes explicit that the gravitational Hamiltonian remains a
constraint.

As a consequence, bulk time evolution is pure gauge: it does not correspond to an
independent physical observable, but rather parametrizes different
representations of the same physical state. Physical information is therefore
encoded in the implementation and resolution of the gravitational constraints
and, in particular, in the structure of the quantum state defined on asymptotic
configurations.

In the asymptotic weak--field region we expand the spatial metric as
\begin{equation}
g_{ij}=\delta_{ij}+h_{ij},\qquad |h_{ij}|\ll 1,
\end{equation}
and organize the Wheeler--DeWitt equation as a perturbative expansion in powers
of \(h_{ij}\) \cite{DeWitt1967,KieferSingh:1991}.
The DeWitt supermetric admits a corresponding expansion,
\begin{equation}
G_{ijkl}(g)
=
G^{(0)}_{ijkl}
+
G^{(1)}_{ijkl}[h]
+
\mathcal O(h^2),
\end{equation}
with leading flat--space expression
\begin{equation}
G^{(0)}_{ijkl}
=
\frac12\Big(\delta_{ik}\delta_{jl}
+\delta_{il}\delta_{jk}
-\delta_{ij}\delta_{kl}\Big).
\end{equation}

At this stage one may formally write a Schr\"odinger--type
equation for the gravitational wave functional,
\begin{equation}
i\,\frac{\partial}{\partial \tau}\,\Psi[h;\tau]
=
\widehat H_{\rm tot}\,\Psi[h;\tau],
\label{22}
\end{equation}
where $\tau$ parametrizes a chosen foliation.

However, Eq.~\eqref{22} does not yet describe operational
quantum evolution in the asymptotic sense, since the
relevant infrared degrees of freedom accessible to an
external observer have not yet been identified.
In the following we implement a Born--Oppenheimer
separation between slow asymptotic configurations and
fast bulk fluctuations, which will allow us to isolate
the genuinely infrared sector and promote the
Regge--Teitelboim boundary charge to an effective
dynamical generator.

At leading order in the asymptotic expansion the dominant bulk contribution is
determined by the background supermetric $G^{(0)}_{ijkl}$.
Corrections encoded in $G^{(1)}_{ijkl}[h]$ are suppressed by an additional power
of the metric perturbation and consistently belong to subleading order.
Accordingly, one may evaluate the supermetric on the flat background,
\[
G_{ijkl}(g)\;\longrightarrow\;G^{(0)}_{ijkl}(\delta),
\]
with metric--dependent corrections relegated to higher--order terms.

We now decompose the asymptotic metric perturbation into slow and fast
components,
\[
h_{ij}=h_{1\,ij}+h_{2\,ij},
\]
where $h_1$ parametrizes infrared collective configurations and $h_2$
represents bulk fluctuations. This separation is not based on a sharp momentum
cutoff but defines an adiabatic decomposition on the space of admissible
asymptotic metrics, in direct analogy with the Born--Oppenheimer construction
\cite{BroutVenturi:1989,KieferSingh:1991}.

The functional Schr\"odinger equation then admits a factorization of the form
\begin{equation}
\Psi[h_1,h_2;\tau]
=
\Phi_0[h_2;h_1]\,
\chi[h_1;\tau],
\label{ans1}
\end{equation}
where \(\Phi_0[h_2;h_1]\) is the instantaneous ground state of the fast sector
for fixed \(h_1\).

Adiabatic transport along a path in the space of slow configurations induces a
functional Berry connection,
\begin{equation}
\mathcal A^{ij}(x;h_1)
=
i\,\big\langle
\Phi_0[h_1]\big|
\frac{\delta}{\delta h_{1\,ij}(x)}
\Phi_0[h_1]
\big\rangle,
\end{equation}
capturing the geometric response of ultraviolet degrees of freedom to slow
deformations of the asymptotic configuration.

Projecting onto the instantaneous ground state yields an effective equation for
\(\chi[h_1;\tau]\), in which functional derivatives are replaced by covariant
derivatives,
\begin{equation}
-i\frac{\delta}{\delta h_{1\,ij}}
\;\longrightarrow\;
-i\frac{\delta}{\delta h_{1\,ij}}
-\,\mathcal A^{ij}.
\end{equation}

In asymptotically flat gravity, however, the generator of physical evolution is
not given by the bulk Hamiltonian constraint alone. A well--defined canonical
generator requires the addition of a Regge--Teitelboim surface term
\cite{ReggeTeitelboim1974}, so that
\begin{equation}
H_{\rm tot}[N_\perp,N^i]
=
\int_\Sigma d^3x
\left(
N_\perp\,\mathcal H_\perp
+
N^i\,\mathcal H_i
\right)
+
Q_{\rm RT}[N_\perp,N^i].
\end{equation}
On the constraint surface, the bulk term vanishes and the physical Hamiltonian
reduces to the boundary charge,
\begin{equation}
H_{\rm phys}=Q_{\rm RT}.
\end{equation}

Once an asymptotic proper--time gauge is chosen, the quantum evolution of
infrared states is therefore governed by the corresponding boundary operator,
\begin{equation}
i\,\partial_s \Psi_{\rm asy}
=
\widehat{Q}_{\rm RT}\,\Psi_{\rm asy}.
\end{equation}

Adiabatic transport in the space of asymptotic configurations,
parametrized by Regge--Teitelboim charges
$\Lambda^A=\{E_{\rm RT},P_i,J_{ij},\ldots\}$,
then induces a Berry connection
\begin{equation}
\mathcal A=\mathcal A_A(\Lambda)\,d\Lambda^A,
\qquad
\mathcal A_A(\Lambda)=
i\langle 0;\Lambda|\partial_{\Lambda^A}|0;\Lambda\rangle,
\end{equation}
whose holonomy along a closed loop $C$,
\begin{equation}
\mathcal U_C
=
\mathcal P\exp\!\left(
i\oint_C \mathcal A_A(\Lambda)\,d\Lambda^A
\right),
\end{equation}
encodes the geometric phase accumulated by the infrared vacuum.

Single--valuedness of the transported state imposes the global consistency
condition
\begin{equation}
\oint_C \mathcal A_A(\Lambda)\,d\Lambda^A
=
2\pi n,
\qquad n\in\mathbb Z,
\end{equation}
which constrains the admissible asymptotic configurations.
These conditions arise from the topology of the vacuum bundle over the space of
asymptotic charges and should not be interpreted as a local spectral
quantization of bulk observables.

In this sense, the infrared sector of quantum gravity is naturally formulated
in terms of asymptotic states endowed with a geometric structure. Berry
connections and holonomies encode global information about bulk degrees of
freedom that have been coarse--grained, providing an infrared dressing of
physical states consistent with the asymptotic symmetry algebra.

\section{Asymptotic charges and generally covariant theories}

In generally covariant theories, such as general relativity, the canonical
Hamiltonian formulation differs in a crucial way from that of ordinary quantum
field theories. The bulk Hamiltonian density is a linear combination of
first--class constraints,
\begin{equation}
\mathcal H_\perp \approx 0,
\qquad
\mathcal H_i \approx 0,
\end{equation}
which generate normal and tangential diffeomorphisms, respectively
\cite{Dirac1958,ArnowittDeserMisner1962,DeWitt1967}.
As a consequence, the canonical bulk Hamiltonian vanishes weakly on the
constraint surface and does not generate physical evolution on the space of
gauge--invariant states \cite{Kuchar1992,Isham1993}.

This feature reflects the fact that, in gravity, local bulk observables are not
operationally accessible in a gauge--invariant manner. Physical measurements are
therefore naturally associated with asymptotic observers, and the relevant
dynamical information is encoded in boundary data. From this perspective, the
gravitational system accessible to an external observer is not the full bulk
theory, but rather an infrared (IR) sector defined by admissible asymptotic
configurations.

A well--defined generator of evolution in this asymptotic sector is obtained by
supplementing the bulk constraints with appropriate boundary terms. Following
Regge and Teitelboim \cite{ReggeTeitelboim1974,BrownYork1993,WaldZoupas2000}, the
total Hamiltonian takes the form
\begin{equation}
H_{\rm tot}[N_\perp,N^i]
=
\int_\Sigma d^3x\,
\bigl(
N_\perp\,\mathcal H_\perp
+
N^i\,\mathcal H_i
\bigr)
+
Q_{\rm RT}[N_\perp,N^i],
\end{equation}
where the surface contribution $Q_{\rm RT}$ is required for functional
differentiability and encodes the conserved asymptotic charges of the theory
\cite{LeeWald1990}.
On the physical phase space, where the bulk constraints are imposed, the
generator of evolution perceived by asymptotic observers reduces entirely to
this boundary term.

Accordingly, in the infrared sector defined by asymptotic data, the effective
Hamiltonian governing the evolution of physical states is given by
\cite{AshtekarStreubel1981}
\begin{equation}
H_{\rm eff}^{(\infty)}
=
Q_{\rm RT}.
\end{equation}
This relation should not be interpreted as a dynamical identity in the bulk, but
rather as a statement about the generator of evolution in the Hilbert space of
asymptotically resolvable states.

A particularly transparent interpretation emerges upon fixing an asymptotic
proper--time gauge, for instance by choosing
\begin{equation}
N_\perp = 1,
\qquad
N^i = 0,
\end{equation}
so that the evolution parameter $\tau$ labels proper time along asymptotic
observers. In this gauge, the quantum evolution of physical states is governed
by the Schr\"odinger--type equation \cite{BroutVenturi:1989,KieferSingh:1991}
\begin{equation}
i\,\frac{\partial}{\partial \tau}\,\Psi
=
\widehat Q_{\rm RT}\,\Psi.
\label{eq:RT_evolution}
\end{equation}

Equation~\eqref{eq:RT_evolution} therefore constitutes the
Schr\"odinger equation governing the physical evolution of
asymptotic gravitational states. Since the bulk Hamiltonian
vanishes on physical configurations, the Regge--Teitelboim
boundary charge plays the role of the interaction Hamiltonian
with respect to the bulk gauge evolution. 

At this point the parameter $\tau$ acquires a direct
operational meaning: it coincides with the proper time
measured by asymptotic observers. Prior to fixing the
asymptotic proper--time gauge, $\tau$ merely labels a
choice of foliation and does not correspond to a
physical observable. The evolution generated by
$\widehat Q_{\rm RT}$ therefore describes genuine
quantum dynamics with respect to an external reference
frame, rather than gauge motion in the bulk.

Since the bulk Hamiltonian vanishes on physical
configurations, $H_{\rm bulk}\approx0$, the physical
generator of evolution reduces to the Regge--Teitelboim
boundary charge, $H_{\rm phys}=Q_{\rm RT}$.
Equation~\eqref{eq:RT_evolution} is thus already written
in the interaction picture with respect to the bulk
gauge evolution, and directly generates a Dyson
expansion for the infrared evolution operator,
\begin{equation}
U(\tau)
=
\mathcal P
\exp\!\left(
i\int d\tau'\,
\widehat Q_{\rm RT}(\tau')
\right).
\end{equation}
The Born--Oppenheimer approximation is not required to
define this interaction picture, but rather ensures the
adiabatic decoupling between slow asymptotic
configurations and fast bulk fluctuations, endowing the
asymptotic Hilbert space with a Berry connection that
characterizes the dressed infrared gravitational
sector.

\paragraph{Remark: dynamical versus geometric phase in asymptotic gravity.}

The asymptotic evolution generated by Eq.~(\ref{eq:RT_evolution}) admits a
natural adiabatic interpretation. In Berry--type systems
\cite{Simon1983,Berry1984,ShapereWilczek1989}, the evolution of slow degrees of
freedom factorizes into a dynamical phase, determined by the instantaneous
energy, and a geometric contribution associated with parallel transport in
parameter space.

In the present context, the Regge--Teitelboim charge
$\widehat Q_{\rm RT}$ plays the role of the physical
generator of asymptotic evolution in the infrared
sector. This identification does not refer to the
microscopic Hamiltonian of the bulk gravitational
theory; rather, after imposing the constraints and
restricting to asymptotically accessible observables,
the generator of evolution in the asymptotic Hilbert
space coincides with the boundary charge.

Within an adiabatic Born--Oppenheimer treatment,
the effective infrared configuration space may be
parametrized by a finite set of Regge--Teitelboim
charges,
\begin{equation}
\Lambda^A=\{E_{\rm RT},P_i,J_{ij},\ldots\},
\end{equation}
which encode the slow gravitational data accessible
to asymptotic observers. Eliminating fast bulk
fluctuations induces an additional geometric
structure on this space: a Berry connection
associated with parallel transport in the space
of Regge--Teitelboim charges.

Restricting the functional Berry connection to
this subspace, one may write
\begin{equation}
\mathcal A=\mathcal A_A(\Lambda)\,d\Lambda^A,
\qquad
\mathcal A_A(\Lambda)=
i\langle 0;\Lambda|\partial_{\Lambda^A}|0;\Lambda\rangle .
\end{equation}
The holonomy associated with an adiabatic loop
$C:\lambda\mapsto\Lambda^A(\lambda)$ is then
\begin{equation}
\mathcal U_C
=
\mathcal P\exp\!\left(
i\oint_C \mathcal A_A(\Lambda)\,d\Lambda^A
\right),
\end{equation}
which characterizes the geometric dressing of
infrared states under adiabatic transport in
the space of asymptotic data.

The asymptotic evolution operator therefore
remains generated by $\widehat Q_{\rm RT}$,
\begin{equation}
\Psi(\tau)
=
\mathcal T\exp\!\left(
-i\int^\tau d\tau'\,
\widehat Q_{\rm RT}(\tau')
\right)\,
\Psi(0),
\end{equation}
while the Berry phase provides a geometric
structure on the asymptotic Hilbert space
induced by bulk degrees of freedom that have
been coarse--grained.

\section{Quantum Gravity and Axion Matter}

In this section we consider a particular coupling of matter to quantum gravity,
focusing on an axion--like pseudoscalar field coupled to the gravitational
Pontryagin density. Specifically, we study the interaction
\begin{equation}
g\,\varphi\,\tilde R_{\mu\nu\rho\lambda} R^{\mu\nu\rho\lambda}
\;=\;
g\,\varphi\,\mathcal P,
\qquad
\mathcal P := \tilde R_{\mu\nu\rho\lambda} R^{\mu\nu\rho\lambda},
\end{equation}
where \(\varphi\) plays the role of a \emph{gravitational axion}, \(g\) has
dimensions of \((\text{mass})^{-1}\), and \(\mathcal P\) denotes the
Pontryagin density \cite{JackiwPi2003}.
As we show below, this topological coupling gives rise to an additional
boundary contribution to the Regge--Teitelboim charge \(Q_{\rm RT}\).

The corresponding action term
\begin{equation}
S_{\varphi\mathcal P}
=
g\int_M d^4x\,\sqrt{-g}\;\varphi\,\mathcal P
\label{eq:axionPontryagin}
\end{equation}
may be rewritten using the identity \cite{Eguchi1980}
\begin{equation}
\sqrt{-g}\,\mathcal P
=
\partial_\mu\!\left(\sqrt{-g}\,K^\mu_{\rm CS}\right),
\label{eq:PontryaginTotalDer}
\end{equation}
where \(K^\mu_{\rm CS}\) denotes the gravitational Chern--Simons current.
Substituting this expression yields
\begin{equation}
\int_M d^4x\,\sqrt{-g}\;\varphi\,\mathcal P
=
\int_M \partial_\mu\!\left(\sqrt{-g}\,\varphi K^\mu_{\rm CS}\right)
-
\int_M d^4x\,\sqrt{-g}\;(\nabla_\mu\varphi)\,K^\mu_{\rm CS}.
\label{eq:axionSplit}
\end{equation}
The first term corresponds to a pure boundary contribution \cite{Tachikawa2007},
\begin{equation}
S_{\rm bdy}
=
g\int_{\partial M} d\Sigma_\mu\;
\varphi\,K^\mu_{\rm CS},
\label{eq:axionBoundary}
\end{equation}
while the second contributes only to bulk dynamics through derivatives of
\(\varphi\). In the infrared regime of interest, where \(\varphi\) varies
slowly at spatial infinity, this bulk term is subleading and does not affect
the asymptotic dynamics.

For minimally coupled scalar fields, such as the kinetic term
\begin{equation}
S_{\rm kin}[\varphi]
=
-\frac12\int_M d^4x\,\sqrt{-g}\;(\nabla\varphi)^2,
\label{eq:axionKinetic}
\end{equation}
the Regge--Teitelboim prescription introduces no additional surface
contributions under standard asymptotically flat fall--off conditions.
Consequently, the kinetic term contributes only to the bulk Hamiltonian and
does not modify asymptotic charges.

In contrast, the boundary term~\eqref{eq:axionBoundary} survives at infinity,
leading to the total generator
\begin{equation}
H_{\rm tot}
=
\int_\Sigma d^3x\,
\bigl(
N_\perp \mathcal H_\perp
+
N^i \mathcal H_i
\bigr)
+
Q_{\rm RT}^{\rm grav}
+
Q_{\rm RT}^{(\varphi)},
\label{eq:HtotAxion}
\end{equation}
with axionic contribution
\begin{equation}
Q_{\rm RT}^{(\varphi)}
=
g\int_{S^2_\infty} dS_i\;
\varphi\,K^i_{\rm CS}.
\label{eq:QRTaxion}
\end{equation}

\paragraph{Infrared factorization and topological dressing.}

Separating the axion field into its asymptotic zero mode and subleading
fluctuations,
\begin{equation}
\varphi(x)
=
\varphi_\infty(\tau)
+
\delta\varphi(x),
\qquad
\delta\varphi \to 0 \;\text{at}\; S^2_\infty,
\end{equation}
the surface charge reduces in the infrared regime to
\begin{equation}
Q_{\rm RT}^{(\varphi)}
=
g\,\varphi_\infty\,
\mathcal I_{\rm CS}[h],
\end{equation}
where $\mathcal I_{\rm CS}[h]$ denotes the Chern--Simons functional
associated with the asymptotic geometry.

In the Schr\"odinger functional representation, this contribution
acts multiplicatively and modifies the phase of asymptotic states,
\begin{equation}
\Psi[h,\varphi;\tau]
\;\simeq\;
\exp\!\Big(i\,g\,\varphi_\infty\,\mathcal I_{\rm CS}[h]\Big)\,
\Psi_{\rm red}[h;\tau],
\label{eq:axionDressingPsi}
\end{equation}
where $\Psi_{\rm red}$ denotes the reduced wave functional obtained
after projecting onto the instantaneous ground state of the fast sector,
in analogy with the Born--Oppenheimer construction introduced previously.

Since the Chern--Simons functional is defined only modulo discrete shifts
under large transformations of the asymptotic configuration
\cite{Witten1989},
\begin{equation}
\mathcal I_{\rm CS}\rightarrow \mathcal I_{\rm CS}+2\pi n,
\qquad n\in\mathbb Z,
\end{equation}
the dressing phase in Eq.~\eqref{eq:axionDressingPsi} is admissible only if
\begin{equation}
g\,\varphi_\infty \in \mathbb Z,
\end{equation}
ensuring that the asymptotic quantum state remains single--valued.
Equivalently, $\varphi_\infty$ behaves as a periodic variable with
period $2\pi/g$.

When $\varphi_\infty$ is transported adiabatically in the space of
asymptotic data, the accumulated boundary phase defines an infrared
Berry holonomy compatible with the global consistency of the quantum
state.

In summary, while the bulk equations of motion are unaffected at leading
infrared order, the topological structure of the Pontryagin term gives rise
to a nontrivial contribution to the Regge--Teitelboim generator through a
surface term. Within the Born--Oppenheimer framework, the asymptotic axion
mode $\varphi_\infty$ plays the role of a slow collective variable, and the
induced boundary term manifests itself as a universal infrared dressing of
asymptotic quantum states.

\section{Cosmological application: Axion couplings}

Although the axionic coupling considered here involves the gravitational
Pontryagin density and manifests itself as a modification of the
Regge--Teitelboim charge, the underlying mechanism is more general: when the
pseudoscalar field instead couples to the electromagnetic Pontryagin density,
the same topological structure reappears in a cosmological setting, where the
slow time evolution of the axion field replaces the role of spatial infinity and
renders the associated geometric phase observationally accessible through
electromagnetic propagation.

A recurrent difficulty in identifying observable consequences of topological
infrared effects in gravity is that they are naturally encoded in boundary
charges or global phases, rather than in local excitations. In asymptotically
flat spacetimes such effects are associated with Regge--Teitelboim charges and
holonomies on the space of asymptotic configurations, whose physical
manifestation is often indirect. Cosmology, however, provides a qualitatively
different setting in which the same type of topological structure becomes
operationally accessible.

Consider a pseudoscalar field $\varphi$ coupled to electromagnetism through the
topological interaction \cite{Sikivie1983}
\begin{equation}
S_{\varphi F\tilde F}
=
\frac{g}{4}\int d^4x\,\sqrt{-g}\;
\varphi\,F_{\mu\nu}\tilde F^{\mu\nu}.
\label{eq:axionEM}
\end{equation}
In a homogeneous and isotropic cosmological background,
\begin{equation}
ds^2=-dt^2+a^2(t)d\vec x^{\,2},
\qquad
\sqrt{-g}=a^3(t),
\end{equation}
 the axion field may be consistently taken to depend
only on cosmic time, $\varphi=\varphi(t)$ \cite{HarariSikivie1992}.
 In this case one finds
\begin{equation}
F_{\mu\nu}\tilde F^{\mu\nu}
=
-\frac{4}{a^3(t)}\,
\vec E\cdot\vec B,
\end{equation}
so that the metric determinant cancels explicitly in the
topological interaction \eqref{eq:axionEM} \cite{CarrollFieldJackiw1990},

\begin{equation}
S_{\varphi F\tilde F}
=
-g\int dt\,d^3x\;
\varphi(t)\,
\vec E\cdot\vec B.
\end{equation}
Integrating by parts yields an effective Chern--Simons
coupling with a time--dependent coefficient,
\begin{equation}
S_{\rm CS}^{\rm eff}
=
g\int dt\,\dot\varphi(t)
\int d^3x\,\vec A\cdot\vec B,
\end{equation}
which is independent of the cosmological scale factor.

In this case, the
interaction \eqref{eq:axionEM} reduces, up to total derivatives, to an effective
Chern--Simons term for the electromagnetic field,
\begin{equation}
S_{\rm CS}^{\rm eff}
\;\sim\;
g\int dt\,\dot\varphi(t)
\int d^3x\,\vec A\cdot\vec B,
\end{equation}
with a time--dependent coefficient controlled by the slow cosmological evolution
of $\varphi$.

The physical implication of this term is simple and robust. Electromagnetic
modes of opposite helicity propagate with slightly different phases, so that a
linearly polarized wave experiences a rotation of its polarization plane during
cosmic propagation. The accumulated rotation angle is given by
\begin{equation}
\Delta\alpha
=
\frac{g}{2}\bigl[\varphi(t_0)-\varphi(t_{\rm em})\bigr],
\label{eq:cosmicRotation}
\end{equation}
where $t_{\rm em}$ is the emission time and $t_0$ denotes the present epoch \cite{CarrollFieldJackiw1990,Kamionkowski2010}.

Several features make this effect particularly transparent. First, it does not
rely on the existence of new propagating degrees of freedom or on modified
dispersion relations; it is entirely controlled by a topological interaction.
Second, the result \eqref{eq:cosmicRotation} is insensitive to the detailed
dynamics of the electromagnetic source and depends only on the net change of the
axion field between emission and observation. Finally, the effect is parity--odd
and frequency--independent to leading order, which sharply distinguishes it
from conventional plasma or refractive effects.

From the infrared viewpoint, the cosmological axion field plays the role of a
slowly varying global parameter. The rotation angle \eqref{eq:cosmicRotation}
can be interpreted as the temporal realization of a geometric phase \cite{Berry1984,Bliokh2015}: 
 the
electromagnetic field is transported adiabatically through a one--parameter
family of configurations labeled by $\varphi(t)$, and the observed polarization
rotation is the holonomy accumulated along this trajectory. In this sense,
cosmology converts an abstract boundary holonomy into a directly measurable
effect.

This mechanism provides a clear example of how topological infrared structures,
which may appear hidden or formal in asymptotically flat gravity, acquire
concrete observational meaning in a cosmological context. The role played by
polarization rotation here is closely analogous to the role of the blackbody
spectrum in infrared QED: in both cases a universal observable probes global
infrared properties of the quantum state rather than local dynamics.

\subsection{Infrared quantum structure and adiabatic transport}

The cosmological axionic coupling does not merely induce a classical rotation
of polarization. From the infrared quantum viewpoint it defines a nontrivial
structure on the Hilbert space of electromagnetic states.

In a homogeneous background $\varphi=\varphi(t)$ the interaction
\eqref{eq:axionEM} modifies the canonical momentum conjugate to the vector
potential,
\begin{equation}
\Pi^i
=
-\,E^i
+
g\,\varphi(t)\, B^i ,
\end{equation}
so that the Hamiltonian becomes
\begin{equation}
H(t)
=
\frac12 \int d^3x
\left[
\bigl(E^i - g\,\varphi(t) B^i\bigr)^2
+
B^i B^i
\right].
\end{equation}
The time dependence of $\varphi$ therefore induces a slow deformation of the
electromagnetic phase space.

For each fixed value of $\varphi$ one may define instantaneous helicity
eigenmodes $|k,\pm;\varphi\rangle$. In the adiabatic regime
$\dot\varphi/H \ll \omega_k$, where $\omega_k$ denotes the photon frequency,
these modes evolve according to
\begin{equation}
i\frac{d}{dt} |k,\pm;t\rangle
=
H(t)\,|k,\pm;t\rangle ,
\end{equation}
and acquire both a dynamical and a geometric phase.

The Berry connection associated with the slow parameter $\varphi$ is
\begin{equation}
\mathcal A_\pm(\varphi)
=
i \langle k,\pm;\varphi |
\partial_\varphi
| k,\pm;\varphi \rangle ,
\end{equation}
and a straightforward computation yields
\begin{equation}
\mathcal A_\pm(\varphi)
=
\pm \frac{g}{2}.
\end{equation}
The geometric phase accumulated between emission and observation is therefore
\begin{equation}
\gamma_\pm
=
\int_{\varphi(t_{\rm em})}^{\varphi(t_0)}
\mathcal A_\pm(\varphi)\, d\varphi
=
\pm \frac{g}{2}
\bigl[\varphi(t_0)-\varphi(t_{\rm em})\bigr],
\end{equation}
which reproduces the rotation angle \eqref{eq:cosmicRotation} as the relative
Berry phase between opposite helicities.

From this perspective the cosmological birefringence is not simply a classical
effect but the infrared holonomy of the photon Hilbert bundle over the
one--dimensional parameter space defined by $\varphi(t)$. The slow cosmological
evolution plays the role of the adiabatic trajectory, and the observable
polarization rotation corresponds to the difference of Berry phases for the two
helicity sectors.

Several points are worth emphasizing.

First, the effect survives in the strict infrared limit $k\to 0$, where local
propagation effects become negligible. The observable is entirely controlled by
the global displacement in field space, not by ultraviolet details.

Second, the result depends only on the path in the parameter space of
$\varphi$, not on the microscopic structure of the source. In this sense it is
a genuine infrared observable.

Finally, this construction parallels the role played by Regge--Teitelboim
charges in asymptotically flat gravity. There, the infrared structure is
encoded in boundary generators and associated holonomies; here, the cosmological
background replaces spatial infinity and the Berry holonomy becomes directly
measurable through photon polarization.

Cosmology thus provides a concrete realization of an infrared quantum
holonomy, making manifest that topological structures in gauge theory and
gravity can acquire observational meaning when the slow parameter is promoted
from a boundary label to a dynamical cosmological field.

From the Berry perspective, the polarization rotation
$\Delta\alpha$ corresponds to the holonomy of the photon
Hilbert bundle over the one--dimensional parameter space
defined by $\varphi(t)$. Requiring the transported
quantum state to remain single--valued along closed
trajectories in this space imposes the consistency
condition
\begin{equation}
g\,\Delta\varphi
\in
2\pi\mathbb Z,
\end{equation}
which can be viewed as an infrared quantization of the
axion--photon coupling from the global geometry of the
asymptotic Hilbert space.

\section{Three--dimensional Asymptotic Quantum Gravity}

Before specializing to black hole geometries, it is useful to recall that the
appropriate starting point for gravity in $2+1$ dimensions with negative
cosmological constant is the asymptotic analysis of Brown and Henneaux
\cite{BrownHenneaux1986}. Their result establishes that the physical content
of the theory is encoded in boundary degrees of freedom, independently of the
existence of horizons or particular classical solutions.

\subsection{Brown--Henneaux asymptotic structure}

Consider Einstein gravity in $2+1$ dimensions with cosmological constant
$\Lambda=-1/\ell^2$ and impose asymptotically $\mathrm{AdS}_3$ boundary
conditions. Brown and Henneaux showed that the allowed asymptotic
diffeomorphisms form two copies of the Virasoro algebra,
\begin{equation}
\mathrm{Vir}_L \times \mathrm{Vir}_R,
\end{equation}
with classical central charge
\begin{equation}
c = \frac{3\ell}{2G}.
\end{equation}

Although the BTZ black hole is a classical solution of the bulk Einstein
equations \cite{BTZ1992}, the physical quantum theory under consideration is
the asymptotic boundary theory. In asymptotically $\mathrm{AdS}_3$ gravity the
bulk carries no local propagating degrees of freedom
\cite{DeserJackiw1984,Witten1988}, and genuine quantum dynamics is encoded in
boundary operators and states. The associated conserved quantities are surface
charges defined at infinity \cite{BrownHenneaux1986}. On the constraint
surface, the bulk Hamiltonian vanishes and the full generator of dynamics
reduces to boundary terms. Thus, for an asymptotic observer, the theory is
governed by global charges and their algebra, rather than by local bulk
excitations.

\subsection{Asymptotic charges as slow collective coordinates}

The Brown--Henneaux analysis naturally defines a finite--dimensional space of
asymptotic configurations coordinatized by the conserved charges associated
with boundary diffeomorphisms. In particular, the zero modes
$(L_0,\bar L_0)$ of the Virasoro generators play the role of global collective
variables.

Within the adiabatic framework developed in this work, these asymptotic charges
may be identified as slow variables, while all remaining degrees of
freedom—including bulk fluctuations and boundary excitations with nonzero
modes—are treated as fast variables that adjust instantaneously to a given
choice of $(L_0,\bar L_0)$. This separation is not based on local energy
scales, but on the global rigidity of the asymptotic data in configuration
space.

Let $|0;\Lambda\rangle$ denote the instantaneous vacuum of the fast sector for
fixed asymptotic charges $\Lambda^A$. Adiabatic transport in the space of
charges induces a Berry connection,
\begin{equation}
\mathcal{A}
=
\mathcal{A}_A(\Lambda)\, d\Lambda^A,
\qquad
\mathcal{A}_A
=
i\langle 0;\Lambda|\partial_{\Lambda^A}|0;\Lambda\rangle,
\end{equation}
which captures the geometric response of ultraviolet degrees of freedom to
slow deformations of the asymptotic configuration.

\subsection{BTZ black hole as a special orbit in charge space}

The BTZ black hole arises as a particular family of classical solutions within
the general Brown--Henneaux phase space. Its defining parameters—mass $M$ and
angular momentum $J$—are specific combinations \cite{BalasubramanianKraus1999}
of the zero--mode charges,
\begin{equation}
L_0 = \frac{1}{2}(M\ell + J),
\qquad
\bar L_0 = \frac{1}{2}(M\ell - J).
\end{equation}
From this viewpoint, the BTZ geometry does not introduce new degrees of
freedom, but selects a distinguished two--dimensional submanifold of the
asymptotic configuration space.

Consequently, any adiabatic loop in the $(M,J)$ plane may be regarded as a loop
in the space of Brown--Henneaux charges. The associated Berry holonomy,
\begin{equation}
\mathcal{U}_C
=
\mathcal{P}\exp\!\left(
i\oint_C \mathcal{A}_A(\Lambda)\, d\Lambda^A
\right),
\end{equation}
is therefore an asymptotic observable, independent of the presence of a
horizon.

\subsection{Flux quantization and consistency conditions}

Single--valuedness of the infrared vacuum under adiabatic transport imposes a
quantization condition on the Berry flux,
\begin{equation}
\oint_C \mathcal{A}_A(\Lambda)\, d\Lambda^A = 2\pi n,
\qquad n\in\mathbb{Z}.
\end{equation}
For two--parameter families this condition may be written in Stokes form,
\begin{equation}
\int_{\Sigma(C)} d\Lambda^A d\Lambda^B\,
\mathcal{F}_{AB}
=
2\pi n,
\qquad
\mathcal{F}_{AB}
=
\partial_A\mathcal{A}_B-\partial_B\mathcal{A}_A.
\end{equation}

These relations should not be interpreted as a spectral quantization of $M$ or
$J$ viewed as local bulk operators. Rather, they express global consistency
conditions on the admissible asymptotic configurations, arising from the
topology of the vacuum bundle over the space of Brown--Henneaux charges.

\section{Adiabatic transport and Berry phase in JT gravity}

A particularly transparent realization of the adiabatic framework
described above is provided by Jackiw--Teitelboim (JT) gravity
\cite{Teitelboim1983,Jackiw1984}. Unlike higher--dimensional models,
JT gravity admits an exact reduction in which all bulk degrees of
freedom can be integrated out, leaving a boundary quantum mechanical
system that captures the full infrared dynamics.

Starting from the JT action,
\begin{equation}
S
=
\frac{1}{2}\int_M d^2x\,\sqrt{-g}\,
\Phi\,(R+2),
\end{equation}
and imposing asymptotically $\mathrm{AdS}_2$ boundary conditions,
the bulk path integral reduces to the Schwarzian theory for the
boundary reparametrization mode $f(t)$
\cite{Marnelius:1982pf,MaldacenaStanfordYang2016,SaadShenkerStanford2019},
\begin{equation}
S_{\rm Sch}
=
-C\int dt\,\{f(t),t\},
\end{equation}
with $C\sim\Phi_r$ determined by the renormalized boundary value
of the dilaton.

The corresponding Hamiltonian takes the quadratic form
\begin{equation}
H
=
\frac{p^2}{2C},
\label{eq:JTfree}
\end{equation}
whose eigenvalue coincides with the ADM energy of the associated
two--dimensional black hole. The infrared sector of JT gravity is
therefore parametrized entirely by the asymptotic boundary clock
$f(t)$, rather than by local bulk fields.

This structure becomes more transparent upon recalling that, in
conformal gauge, the bulk equations reduce to a Liouville--type
equation for an auxiliary field $\phi(t)$,
\begin{equation}
\ddot\phi(t)=\mu^2 e^{2\phi(t)}.
\end{equation}
The general solution may be written in terms of two independent
functions,
\begin{equation}
e^{2\phi(t)}
=
\frac{f'(t)\,g'(t)}{\big(f(t)-g(t)\big)^2}.
\end{equation}
Asymptotic $\mathrm{AdS}_2$ boundary conditions fix one of these
functions up to global $\mathrm{SL}(2,\mathbb{R})$ transformations,
leaving a single physical reparametrization mode $f(t)$.

Substituting the general solution back into the bulk action and
integrating out the constrained degrees of freedom yields the
Schwarzian functional above. The apparent nonlinearity of the
bulk Liouville description is thus reinterpreted as the nontrivial
geometry of the space of reparametrizations \cite{StanfordWitten2017},
\begin{equation}
\mathrm{Diff}(S^1)/\mathrm{SL}(2,\mathbb{R}),
\end{equation}
on which the genuine infrared degree of freedom evolves according
to the free Hamiltonian~(\ref{eq:JTfree}).

In particular, the ADM energy is not an independent parameter, but
a functional of the boundary clock,
\begin{equation}
E
\;\sim\;
\{f(t),t\},
\end{equation}
so that slow variations of the asymptotic charge correspond to
adiabatic transport in the space of boundary reparametrizations.

From the infrared Born--Oppenheimer viewpoint adopted in this
work, the boundary clock $f(t)$ plays the role of a slow
collective coordinate parametrizing admissible asymptotic
configurations. Bulk excitations and nonzero boundary modes
constitute fast degrees of freedom that adjust instantaneously
to a given trajectory in this space.

Let $|0;f\rangle$ denote the instantaneous ground state of the
fast sector for fixed $f(t)$. Adiabatic transport along a closed
loop
\begin{equation}
C:\;\lambda\mapsto f_\lambda(t)
\end{equation}
in the space of boundary clocks induces a Berry connection,
\begin{equation}
\mathcal A_f
=
i\langle 0;f|\partial_f|0;f\rangle,
\end{equation}
so that the asymptotic quantum state acquires a geometric phase
\begin{equation}
\gamma(C)
=
\oint_C \mathcal A_f\,df.
\end{equation}

The infrared evolution operator therefore factorizes into a
dynamical phase generated by the free Hamiltonian and a
geometric contribution determined by parallel transport in the
space $\mathrm{Diff}(S^1)/\mathrm{SL}(2,\mathbb{R})$. In this
sense, Berry holonomies in JT gravity arise from nontrivial
loops in the space of boundary reparametrizations and encode
global infrared information about bulk degrees of freedom that
have been coarse--grained.

In the present formulation this universal mechanism acquires a
direct quantum--gravitational interpretation: the boundary
reparametrization mode $f(t)$ is promoted to a slow collective
coordinate parametrizing admissible asymptotic configurations,
while bulk excitations constitute fast degrees of freedom that
have been coarse--grained. Adiabatic transport in the resulting
configuration space induces a Berry connection on the asymptotic
Hilbert bundle, so that the Schwarzian dynamics governs not only
the infrared spectrum but also the geometric dressing of
asymptotic states through holonomies in
$\mathrm{Diff}(S^1)/\mathrm{SL}(2,\mathbb{R})$.

It is worth stressing that the emergence of the Schwarzian
derivative as an effective functional governing
reparametrization modes is not unique to the modern JT/SYK
framework. Closely related structures already arise in the
early string/Liouville literature, where the same functional
appears upon fixing reparametrization invariance and
integrating out bulk degrees of freedom (see e.g.
\cite{Marnelius:1982pf}). From this viewpoint, the Schwarzian
sector should be regarded as a universal infrared signature of
reparametrization--invariant dynamics, rather than as a feature
specific to two--dimensional gravity.

\section{Infrared Superselection and Entropic Observables}

The infrared description developed in the previous sections is
constrained by superselection rules arising from the adiabatic
separation between slow (infrared) and fast (ultraviolet) gravitational
modes. In this regime, the effective Hilbert space decomposes into
sectors labelled by geometric data, such as Berry holonomies, and local
dynamical operators cannot induce transitions between them. As a
consequence, standard observables constructed from infrared transition
amplitudes are insensitive to the global structure of these sectors.

This limitation motivates the exploration of observables that remain
accessible within a given superselection sector but are nevertheless
sensitive to ultraviolet--infrared correlations.

A natural candidate is the reduced density matrix obtained by tracing
over ultraviolet bulk fluctuations,
\begin{equation}
\rho_{\rm IR}
=
\operatorname{Tr}_{\rm UV}
|\Psi\rangle\langle\Psi|,
\end{equation}
whose associated von Neumann entropy provides a measure of the
entanglement between infrared configurations and ultraviolet dressing.
Such entropic observables may capture global infrared information that
is otherwise inaccessible to local dynamics, and are therefore
particularly relevant in gravitational systems where gauge constraints
prevent a naive factorization of the physical Hilbert space.

Let us consider a Born--Oppenheimer (adiabatic) decomposition of the
gravitational wavefunctional into slow (infrared) and fast (ultraviolet)
configurations,
\begin{equation}
\Psi[h_1,h_2]
=
\sum_n
\chi_n[h_1]\,
\phi_n[h_2;h_1],
\label{eq:BOdecomposition}
\end{equation}
where $h_1$ denotes infrared gravitational modes and $h_2$ denotes
ultraviolet bulk fluctuations. For each fixed infrared configuration
$h_1$, the fast eigenstates satisfy the parametric eigenvalue problem
\begin{equation}
\widehat Q_{\rm fast}(h_1)\,
\phi_n[h_2;h_1]
=
\varepsilon_n(h_1)\,
\phi_n[h_2;h_1],
\label{eq:fast_eigenproblem}
\end{equation}
with orthonormality condition
\begin{equation}
\int \mathcal D h_2\,
\phi_n[h_2;h_1]\,
\phi_m^*[h_2;h_1]
=
\delta_{nm}.
\label{eq:fast_orthonormality}
\end{equation}

In the adiabatic regime, the dependence of the fast eigenstates on the
slow configuration $h_1$ induces a Berry connection on the space of
infrared fields,
\begin{equation}
\mathcal A_n^{ij}(x;h_1)
=
i
\int \mathcal D h_2\,
\phi_n^*[h_2;h_1]\,
\frac{\delta}{\delta h_{1\,ij}(x)}
\phi_n[h_2;h_1].
\label{eq:Berry_connection}
\end{equation}
Equivalently, the fast eigenstates acquire a geometric phase,
\begin{equation}
\phi_n[h_2;h_1]
=
e^{\,i\theta_n[h_1]}\,
\tilde\phi_n[h_2;h_1],
\label{eq:Berry_phase}
\end{equation}
where the phase functional is given by the line integral of the Berry
connection along a path in configuration space,
\begin{equation}
\theta_n[h_1]
=
\int_\Gamma
\mathcal A_n^{ij}(x;h_1)\,
d h_{1\,ij}(x).
\label{eq:Berry_holonomy}
\end{equation}

Substituting \eqref{eq:Berry_phase} into the decomposition
\eqref{eq:BOdecomposition}, the full wavefunctional becomes
\begin{equation}
\Psi[h_1,h_2]
=
\sum_n
\chi_n[h_1]\,
e^{i\theta_n[h_1]}\,
\tilde\phi_n[h_2;h_1].
\label{eq:BO_with_phase}
\end{equation}

The reduced infrared density matrix is defined by tracing over the
ultraviolet degrees of freedom,
\begin{equation}
\rho_{\rm IR}[h_1,h_1']
=
\int \mathcal D h_2\,
\Psi[h_1,h_2]\,
\Psi^*[h_1',h_2].
\label{eq:RDM_definition}
\end{equation}
Using \eqref{eq:BO_with_phase} and the orthonormality of the fast
eigenstates, one obtains
\begin{equation}
\rho_{\rm IR}[h_1,h_1']
=
\sum_n
\chi_n[h_1]\,
\chi_n^*[h_1']\,
\exp\!\left[
i\bigl(
\theta_n[h_1]
-
\theta_n[h_1']
\bigr)
\right].
\label{eq:RDM_with_phase}
\end{equation}

The reduced density matrix is therefore not simply a classical sum over
infrared amplitudes, but retains non--local phase correlations between
distinct infrared configurations. The phase difference appearing in
\eqref{eq:RDM_with_phase} may be expressed as a holonomy of the Berry
connection along a path connecting $h_1'$ to $h_1$,
\begin{equation}
\theta_n[h_1]
-
\theta_n[h_1']
=
\int_{h_1'}^{h_1}
\mathcal A_n.
\label{eq:holonomy_difference}
\end{equation}

If the Berry curvature associated with $\mathcal A_n$ is quantized,
\begin{equation}
\oint
\mathcal A_n
=
2\pi k,
\qquad
k\in\mathbb Z,
\label{eq:quantization}
\end{equation}
the infrared Hilbert space decomposes into superselection sectors
labelled by $k$. 

Within the infrared theory the corresponding flux operator does not
belong to the algebra of observables preserving the dressed Hilbert
space, as its action would necessarily modify the soft configuration
of the state. As a result, the sector label $k$ becomes a global
property of the dressed state rather than a dynamically measurable
observable.

Operationally, this implies that distinct superselection sectors are
indistinguishable by infrared measurements. The effective infrared
description must therefore be given by a classical mixture over
sectors,
\begin{equation}
\rho_{\rm IR}
=
\sum_k p_k\,\rho_k,
\qquad
\Tr(\rho_k)=1,
\qquad
\sum_k p_k=1,
\end{equation}
so that coarse--graining over unresolved sectors yields an additional
entropy contribution of classical origin,
\begin{equation}
S_{\rm top}
=
-
\sum_k
p_k
\ln p_k,
\label{eq:Stop}
\end{equation}
reflecting the topological structure induced by ultraviolet dressing.

Operationally, this implies that distinct superselection sectors are
indistinguishable by infrared measurements. The effective infrared
description must therefore be given by a classical mixture over
sectors,
\begin{equation}
\rho_{\rm IR}
=
\sum_k p_k\,\rho_k,
\end{equation}
which is analogous to the appearance of superselection sectors as
inequivalent representations of the observable algebra in algebraic
quantum field theory \cite{HaagKastler,Doplicher:1971wk,Doplicher:1973at}, where local
observables cannot induce transitions between sectors carrying
distinct global charges.

\subsection{Induced Berry connection from constrained Born--Oppenheimer reduction}
\label{subsec:Induced_Berry_RT}

We now show that the Berry connection introduced in the previous
sections is not an external assumption, but arises directly from a
Born--Oppenheimer reduction of the constrained gravitational dynamics
once the Regge--Teitelboim boundary term is taken into account.

\smallskip

\noindent
{\bf (i) Total Hamiltonian and asymptotic charges.}

In asymptotically flat gravity, the total Hamiltonian generating
physical time evolution takes the form
\begin{equation}
H_{\rm tot}
=
\int_{\Sigma}
\bigl(
N^\perp \mathcal H_\perp
+
N^i \mathcal H_i
\bigr)
+
Q_{\rm RT}[\Lambda],
\label{eq:Htot_RT}
\end{equation}
where $\mathcal H_\perp$ and $\mathcal H_i$ are the Hamiltonian and
momentum constraints, and $Q_{\rm RT}[\Lambda]$ is the Regge--Teitelboim
boundary term associated with the asymptotic charges
\begin{equation}
\Lambda^A
=
\{E_{\rm ADM},P_i,J_{ij},\ldots\}.
\end{equation}
Physical states satisfy the Wheeler--DeWitt constraints
\begin{equation}
\mathcal H_\perp \Psi = 0,
\qquad
\mathcal H_i \Psi = 0,
\end{equation}
so that the only non--vanishing generator of physical evolution is the
boundary charge $Q_{\rm RT}$.

\smallskip

\noindent
{\bf (ii) Born--Oppenheimer factorization.}

We now separate slow infrared variables (encoded in the asymptotic
charges $\Lambda^A$) from fast bulk gravitational fluctuations $h_2$,
and consider a Born--Oppenheimer factorization of the physical
wavefunctional,
\begin{equation}
\Psi[h_1,h_2]
=
\Phi_0[h_2;\Lambda]\,
\chi[\Lambda],
\label{eq:BO_factorization_RT}
\end{equation}
where $\Phi_0$ is the ground state of the fast sector at fixed values of
the asymptotic charges.

\smallskip

\noindent
{\bf (iii) Projected evolution.}

Physical evolution is generated by the Schrödinger equation
\begin{equation}
i\frac{d}{dt}\Psi
=
H_{\rm tot}\Psi.
\end{equation}
Projecting onto the fast ground state by taking the functional inner
product with $\Phi_0[h_2;\Lambda]$ yields an effective evolution equation
for the slow wavefunctional $\chi[\Lambda]$,
\begin{equation}
i\frac{d}{dt}\chi
=
\Bigl(
Q_{\rm RT}
-
\mathcal A_A(\Lambda)\,\dot{\Lambda}^A
\Bigr)
\chi,
\label{eq:effective_RT}
\end{equation}
where the induced connection is given by
\begin{equation}
\mathcal A_A(\Lambda)
=
i
\int \mathcal D h_2\,
\Phi_0^*[h_2;\Lambda]\,
\frac{\partial}{\partial \Lambda^A}
\Phi_0[h_2;\Lambda].
\label{eq:Berry_RT}
\end{equation}

\smallskip

\noindent
{\bf (iv) Interpretation.}

Equation \eqref{eq:effective_RT} shows that integrating out fast bulk
gravitational fluctuations induces a Berry connection over the space of
asymptotic charges. The physical evolution of the infrared sector is
therefore governed by parallel transport in this charge space,
\begin{equation}
\frac{d}{dt}
\;\to\;
\frac{d}{dt}
-
i\mathcal A_A(\Lambda)\dot{\Lambda}^A,
\end{equation}
which defines a $U(1)$ bundle over the asymptotic phase space.

This result establishes that the geometric phase organizing the infrared
Hilbert space into superselection sectors arises directly from the
constrained gravitational dynamics, rather than from an external
analogy with ordinary Born--Oppenheimer systems.

\subsection{Application to Schwarzschild: asymptotic infrared EFT and entropic sectors}
\label{subsec:Schwarzschild_application}

We now specialize the general infrared superselection construction to a
Schwarzschild black hole as described by an asymptotic observer. No new
formal ingredients are introduced here; we only identify the relevant
background, the infrared degrees of freedom accessible at $S^2_\infty$,
and the interpretation of the sector weights and entropies in this
setting.

\smallskip

\noindent
{\bf (i) Background and mode split.}
We expand the spatial metric around a Schwarzschild saddle,
\begin{equation}
g_{ij}(x)=g^{\rm Sch}_{ij}(x;M,\ldots)+h_{1\,ij}(x)+h_{2\,ij}(x),
\label{eq:BO_split_Sch}
\end{equation}
where $h_1$ collects long--wavelength deformations resolvable from
infinity (including asymptotic memory data), while $h_2$ denotes fast
bulk fluctuations. Tracing over $h_2$ defines an infrared effective
description for $h_1$ at fixed macroscopic charges (e.g.\ $M$),
\begin{equation}
e^{\,iS_{\rm eff}[h_1;g^{\rm Sch}]}
:=
\int \mathcal D h_2\;
e^{\,iS[g^{\rm Sch}+h_1+h_2]}\,.
\label{eq:Seff_def}
\end{equation}
The ``memory'' of the black hole in the infrared description is thus
encoded in the asymptotic charges of the saddle together with the
infrared boundary data carried by $h_1$.

\smallskip

\noindent
{\bf (ii) Asymptotic algebra and dressed sectors.}
The asymptotic observer only accesses gauge--invariant infrared
observables supported at $S^2_\infty$. The adiabatic dependence of the
ultraviolet sector on $h_1$ induces Berry holonomies which, when
quantized as in \eqref{eq:quantization}, label superselection sectors
$k$ of the infrared dressed Hilbert space. The sector label is a global
property of the dressed state and cannot be changed by infrared
dynamical operators.

\smallskip

\noindent
{\bf (iii) Reduced state and entropic observables.}
The reduced infrared density matrix relevant for the asymptotic observer
is the one introduced in \eqref{eq:RDM_definition}, and its sector
decomposition takes the general form
\begin{equation}
\rho_{\rm IR}=\sum_k p_k\,\rho_k.
\label{eq:rhoIR_blocks_Sch}
\end{equation}
The weights $p_k$ quantify how ultraviolet bulk dressing populates the
different infrared holonomy sectors in the Schwarzschild background.

\smallskip

\noindent
{\bf (iv) Schwarzschild entropy with a superselection contribution.}
The von Neumann entropy measured at infinity decomposes as
\begin{equation}
S_{\rm IR}
=
\sum_k p_k\,S(\rho_k)
-
\sum_k p_k\ln p_k
\equiv
\sum_k p_k\,S(\rho_k)
+
S_{\rm top},
\end{equation}
where $S_{\rm top}$ is given in \eqref{eq:Stop}. The first term is the
entropy within a fixed dressed sector, semiclassically dominated by the
same macroscopic Schwarzschild saddle with corrections controlled by
$S_{\rm eff}$ in \eqref{eq:Seff_def}.

\section{Discussion}

The framework developed in this work should be interpreted as an infrared
effective description of the asymptotic gravitational sector accessible to
external observers, in which ultraviolet bulk dynamics manifests itself only
through geometric structures induced on the space of Regge--Teitelboim charges. 
Within this framework, the quantum
dynamics accessible to an asymptotic observer is governed by the
Regge-Teitelboim surface charge, which acts as an effective
Hamiltonian on the infrared Hilbert space. Integrating out fast
bulk fluctuations induces a functional Berry connection on the
space of asymptotic configurations, so that gravitational states
are naturally characterized by geometric phases and holonomies.
This provides a consistent asymptotic description in which
genuinely quantum gravitational information is encoded in the
infrared organization of dressed states rather than in local
bulk observables.
\begingroup
\let\clearpage\relax
\section*{ACKNOWLEDGMENTS}
The authors are grateful to R. Mackenzie and F. Mendez for useful discussions. This research was 
supported by DICYT (USACH), grant number 042531GR\_REG. The work of N.T.A is supported by Agnes Scott College.
\endgroup

\bibliography{refs}

@article{Dirac1958,
    author = "Dirac, Paul A. M.",
    title = "The Theory of gravitation in Hamiltonian form",
    doi = "10.1098/rspa.1958.0142",
    journal = "Proc. Roy. Soc. Lond. A",
    volume = "246",
    pages = "333--343",
    year = "1958"
}

@article{ArnowittDeserMisner1962,
  author = {R. Arnowitt and S. Deser and C. W. Misner},
  title = {The Dynamics of General Relativity},
  journal = {in Gravitation: an Introduction to Current Research},
  year = {1962},
  eprint = {gr-qc/0405109}
}

@article{ReggeTeitelboim1974,
    author = "Regge, Tullio and Teitelboim, Claudio",
    title = "{Role of Surface Integrals in the Hamiltonian Formulation of General Relativity}",
    reportNumber = "Print-74-0988 (IAS,PRINCETON)",
    doi = "10.1016/0003-4916(74)90404-7",
    journal = "Annals Phys.",
    volume = "88",
    pages = "286",
    year = "1974"
}

@article{Bondi1962,
    author = "Bondi, H. and van der Burg, M. G. J. and Metzner, A. W. K.",
    title = "{Gravitational waves in general relativity. 7. Waves from axisymmetric isolated systems}",
    doi = "10.1098/rspa.1962.0161",
    journal = "Proc. Roy. Soc. Lond. A",
    volume = "269",
    pages = "21--52",
    year = "1962"
}

@article{Sachs1962,
    author = "Sachs, R. K.",
    title = "{Gravitational waves in general relativity. 8. Waves in asymptotically flat space-times}",
    doi = "10.1098/rspa.1962.0206",
    journal = "Proc. Roy. Soc. Lond. A",
    volume = "270",
    pages = "103--126",
    year = "1962"
}

@article{Hawking1975,
    author = "Hawking, S. W.",
    editor = "Gibbons, G. W. and Hawking, S. W.",
    title = "{Particle Creation by Black Holes}",
    doi = "10.1007/BF02345020",
    journal = "Commun. Math. Phys.",
    volume = "43",
    pages = "199--220",
    year = "1975",
}

@article{Unruh1976,
    author = "Unruh, W. G.",
    title = "{Notes on black hole evaporation}",
    doi = "10.1103/PhysRevD.14.870",
    journal = "Phys. Rev. D",
    volume = "14",
    pages = "870",
    year = "1976"
}

@article{Chung1965,
    author = "Chung, Victor",
    title = "{Infrared Divergence in Quantum Electrodynamics}",
    doi = "10.1103/PhysRev.140.B1110",
    journal = "Phys. Rev.",
    volume = "140",
    pages = "B1110--B1122",
    year = "1965"
}

@article{Kibble1968,
    author = "Kibble, T. W. B.",
    title = "{Coherent Soft-Photon States and Infrared Divergences. I. Classical Currents}",
    doi = "10.1063/1.1664582",
    journal = "J. Math. Phys.",
    volume = "9",
    number = "2",
    pages = "315--324",
    year = "1968"
}

@article{KulishFaddeev1970,
    author = "Kulish, P. P. and Faddeev, L. D.",
    title = "{Asymptotic conditions and infrared divergences in quantum electrodynamics}",
    reportNumber = "D70-07927",
    doi = "10.1007/BF01066485",
    journal = "Theor. Math. Phys.",
    volume = "4",
    pages = "745",
    year = "1970"
}

@article{Berry1984,
    author = "Berry, Michael V.",
    title = "{Quantal phase factors accompanying adiabatic changes}",
    doi = "10.1098/rspa.1984.0023",
    journal = "Proc. Roy. Soc. Lond. A",
    volume = "392",
    pages = "45--57",
    year = "1984"
}

@book{ShapereWilczek1989,
    editor = "Shapere, Alfred D. and Wilczek, Frank",
    title = "{Geometric Phases in Physics}",
    volume = "5",
    year = "1989"
}

@article{Strominger2014,
    author = "Strominger, Andrew",
    title = "{On BMS Invariance of Gravitational Scattering}",
    eprint = "1312.2229",
    archivePrefix = "arXiv",
    primaryClass = "hep-th",
    doi = "10.1007/JHEP07(2014)152",
    journal = "JHEP",
    volume = "07",
    pages = "152",
    year = "2014"
}

@book{Strominger2017,
    author = "Strominger, Andrew",
    title = "{Lectures on the Infrared Structure of Gravity and Gauge Theory}",
    eprint = "1703.05448",
    archivePrefix = "arXiv",
    primaryClass = "hep-th",
    isbn = "978-0-691-17973-5",
    publisher = "Princeton University Press",
    year = "2018"
}

@inbook{Strominger2018,
    author = "Strominger, Andrew",
    title = "{Black Hole Information Revisited}",
    eprint = "1706.07143",
    archivePrefix = "arXiv",
    primaryClass = "hep-th",
    doi = "10.1142/9789811203961_0010",
    year = "2020"
}

@article{DeWitt1967,
    author = "DeWitt, Bryce S.",
    editor = "Fang, Li-Zhi and Ruffini, R.",
    title = "{Quantum Theory of Gravity. 1. The Canonical Theory}",
    doi = "10.1103/PhysRev.160.1113",
    journal = "Phys. Rev.",
    volume = "160",
    pages = "1113--1148",
    year = "1967"
}

@article{Kuchar1992,
  author = {K. V. Kucha\v{r}},
  title = {Time and Interpretations of Quantum Gravity},
  journal = {Int. J. Mod. Phys. D},
  volume = {20},
  pages = {3},
  year = {2011},
  eprint = {gr-qc/9304012}
}

@article{Isham1993,
  author = {C. J. Isham},
  title = {Canonical Quantum Gravity and the Problem of Time},
  journal = {NATO Sci. Ser. C},
  volume = {409},
  pages = {157},
  year = {1993},
  eprint = {gr-qc/9210011}
}

@article{Rovelli1991,
    author = "Rovelli, Carlo",
    title = "{What Is Observable in Classical and Quantum Gravity?}",
    reportNumber = "PITT-90-10",
    doi = "10.1088/0264-9381/8/2/011",
    journal = "Class. Quant. Grav.",
    volume = "8",
    pages = "297--316",
    year = "1991"
}

@article{Weinberg1965,
    author = "Weinberg, Steven",
    title = "{Infrared photons and gravitons}",
    doi = "10.1103/PhysRev.140.B516",
    journal = "Phys. Rev.",
    volume = "140",
    pages = "B516--B524",
    year = "1965"
}

@article{Yennie1961,
    author = "Yennie, D. R. and Frautschi, Steven C. and Suura, H.",
    title = "{The infrared divergence phenomena and high-energy processes}",
    doi = "10.1016/0003-4916(61)90151-8",
    journal = "Annals Phys.",
    volume = "13",
    pages = "379--452",
    year = "1961"
}

@article{Buchholz1986,
    author = "Buchholz, Detlev",
    title = "{Gauss' Law and the Infraparticle Problem}",
    reportNumber = "DESY-86-035",
    doi = "10.1016/0370-2693(86)91110-X",
    journal = "Phys. Lett. B",
    volume = "174",
    pages = "331--334",
    year = "1986"
}

@article{BroutVenturi:1989,
    author = "Brout, R. and Venturi, Giovanni",
    title = "{Time in Semiclassical Gravity}",
    reportNumber = "ULB-TH-88/09",
    doi = "10.1103/PhysRevD.39.2436",
    journal = "Phys. Rev. D",
    volume = "39",
    pages = "2436",
    year = "1989"
}

@article{KieferSingh:1991,
    author = "Kiefer, Claus and Singh, Tejinder P.",
    title = "{Quantum gravitational corrections to the functional Schrodinger equation}",
    reportNumber = "IC-90-312",
    doi = "10.1103/PhysRevD.44.1067",
    journal = "Phys. Rev. D",
    volume = "44",
    pages = "1067--1076",
    year = "1991"
}

@article{BrownYork1993,
    author = "Brown, J. David and York, Jr., James W.",
    title = "{Quasilocal energy and conserved charges derived from the gravitational action}",
    eprint = "gr-qc/9209012",
    archivePrefix = "arXiv",
    reportNumber = "IFP-423-UNC, TAR-009-UNC",
    doi = "10.1103/PhysRevD.47.1407",
    journal = "Phys. Rev. D",
    volume = "47",
    pages = "1407--1419",
    year = "1993"
}

@article{WaldZoupas2000,
    author = "Wald, Robert M. and Zoupas, Andreas",
    title = "{A General definition of 'conserved quantities' in general relativity and other theories of gravity}",
    eprint = "gr-qc/9911095",
    archivePrefix = "arXiv",
    doi = "10.1103/PhysRevD.61.084027",
    journal = "Phys. Rev. D",
    volume = "61",
    pages = "084027",
    year = "2000"
}

@article{LeeWald1990,
    author = "Lee, J. and Wald, Robert M.",
    title = "{Local symmetries and constraints}",
    doi = "10.1063/1.528801",
    journal = "J. Math. Phys.",
    volume = "31",
    pages = "725--743",
    year = "1990"
}

@article{AshtekarStreubel1981,
    author = "Ashtekar, A. and Streubel, M.",
    title = "{Symplectic Geometry of Radiative Modes and Conserved Quantities at Null Infinity}",
    doi = "10.1098/rspa.1981.0109",
    journal = "Proc. Roy. Soc. Lond. A",
    volume = "376",
    pages = "585--607",
    year = "1981"
}

@article{Simon1983,
    author = "Simon, Barry",
    title = "{Holonomy, the quantum adiabatic theorem, and Berry's phase}",
    doi = "10.1103/PhysRevLett.51.2167",
    journal = "Phys. Rev. Lett.",
    volume = "51",
    pages = "2167--2170",
    year = "1983"
}

@article{JackiwPi2003,
    author = "Jackiw, R. and Pi, S. Y.",
    title = "{Chern-Simons modification of general relativity}",
    eprint = "gr-qc/0308071",
    archivePrefix = "arXiv",
    reportNumber = "MIT-CTP-3409, BUHEP-03-18",
    doi = "10.1103/PhysRevD.68.104012",
    journal = "Phys. Rev. D",
    volume = "68",
    pages = "104012",
    year = "2003"
}

@article{Eguchi1980,
    author = "Eguchi, Tohru and Gilkey, Peter B. and Hanson, Andrew J.",
    title = "{Gravitation, Gauge Theories and Differential Geometry}",
    reportNumber = "PRINT-80-0182 (LBL,BERKELEY)",
    doi = "10.1016/0370-1573(80)90130-1",
    journal = "Phys. Rept.",
    volume = "66",
    pages = "213",
    year = "1980"
}

@article{Tachikawa2007,
    author = "Tachikawa, Yuji",
    title = "{Black hole entropy in the presence of Chern-Simons terms}",
    eprint = "hep-th/0611141",
    archivePrefix = "arXiv",
    doi = "10.1088/0264-9381/24/3/014",
    journal = "Class. Quant. Grav.",
    volume = "24",
    pages = "737--744",
    year = "2007"
}

@article{Witten1989,
    author = "Witten, Edward",
    editor = "Mitra, Asoke N.",
    title = "{Quantum Field Theory and the Jones Polynomial}",
    reportNumber = "IASSNS-HEP-88-33",
    doi = "10.1007/BF01217730",
    journal = "Commun. Math. Phys.",
    volume = "121",
    pages = "351--399",
    year = "1989"
}

@article{Sikivie1983,
    author = "Sikivie, P.",
    editor = "Srednicki, M. A.",
    title = "{Experimental Tests of the Invisible Axion}",
    reportNumber = "PRINT-83-0597 (FLORIDA), UF-TP-83-13",
    doi = "10.1103/PhysRevLett.51.1415",
    journal = "Phys. Rev. Lett.",
    volume = "51",
    pages = "1415--1417",
    year = "1983",
    note = "[Erratum: Phys.Rev.Lett. 52, 695 (1984)]"
}

@article{CarrollFieldJackiw1990,
    author = "Carroll, Sean M. and Field, George B. and Jackiw, Roman",
    title = "{Limits on a Lorentz and Parity Violating Modification of Electrodynamics}",
    reportNumber = "MIT-CTP-1782",
    doi = "10.1103/PhysRevD.41.1231",
    journal = "Phys. Rev. D",
    volume = "41",
    pages = "1231",
    year = "1990"
}

@article{HarariSikivie1992,
    author = "Harari, Diego and Sikivie, Pierre",
    title = "{Effects of a Nambu-Goldstone boson on the polarization of radio galaxies and the cosmic microwave background}",
    reportNumber = "UFIFT-HEP-92-9",
    doi = "10.1016/0370-2693(92)91363-E",
    journal = "Phys. Lett. B",
    volume = "289",
    pages = "67--72",
    year = "1992"
}

@article{Kamionkowski2010,
    author = "Kamionkowski, Marc",
    title = "{How to De-Rotate the Cosmic Microwave Background Polarization}",
    eprint = "0810.1286",
    archivePrefix = "arXiv",
    primaryClass = "astro-ph",
    doi = "10.1103/PhysRevLett.102.111302",
    journal = "Phys. Rev. Lett.",
    volume = "102",
    pages = "111302",
    year = "2009"
}

@article{Bliokh2015,
    author = "Bliokh, Konstantin Y. and Nori, Franco",
    title = "{Transverse and longitudinal angular momenta of light}",
    eprint = "1504.03113",
    archivePrefix = "arXiv",
    primaryClass = "physics.optics",
    doi = "10.1016/j.physrep.2015.06.003",
    journal = "Phys. Rept.",
    volume = "592",
    pages = "1--38",
    year = "2015"
}

@article{BrownHenneaux1986,
    author = "Brown, J. David and Henneaux, M.",
    title = "{Central Charges in the Canonical Realization of Asymptotic Symmetries: An Example from Three-Dimensional Gravity}",
    doi = "10.1007/BF01211590",
    journal = "Commun. Math. Phys.",
    volume = "104",
    pages = "207--226",
    year = "1986"
}

@article{DeserJackiw1984,
    author = "Deser, Stanley and Jackiw, R.",
    title = "{Three-Dimensional Cosmological Gravity: Dynamics of Constant Curvature}",
    reportNumber = "MIT-CTP-1106",
    doi = "10.1016/0003-4916(84)90025-3",
    journal = "Annals Phys.",
    volume = "153",
    pages = "405--416",
    year = "1984"
}

@article{BTZ1992,
    author = "Banados, Maximo and Teitelboim, Claudio and Zanelli, Jorge",
    title = "{The Black hole in three-dimensional space-time}",
    eprint = "hep-th/9204099",
    archivePrefix = "arXiv",
    reportNumber = "PRINT-92-0151 (CHILE), IASSNS-HEP-92-29",
    doi = "10.1103/PhysRevLett.69.1849",
    journal = "Phys. Rev. Lett.",
    volume = "69",
    pages = "1849--1851",
    year = "1992"
}

@article{BalasubramanianKraus1999,
    author = "Balasubramanian, Vijay and Kraus, Per",
    title = "{A Stress tensor for Anti-de Sitter gravity}",
    eprint = "hep-th/9902121",
    archivePrefix = "arXiv",
    reportNumber = "HUTP-99-A002, EFI-99-6, NSF-ITP-98-132",
    doi = "10.1007/s002200050764",
    journal = "Commun. Math. Phys.",
    volume = "208",
    pages = "413--428",
    year = "1999"
}

@article{Witten1988,
    author = "Witten, Edward",
    title = "{(2+1)-Dimensional Gravity as an Exactly Soluble System}",
    reportNumber = "IASSNS-HEP-88-32",
    doi = "10.1016/0550-3213(88)90143-5",
    journal = "Nucl. Phys. B",
    volume = "311",
    pages = "46",
    year = "1988"
}

@article{Jackiw1984,
    author = "Jackiw, R.",
    editor = "Baier, R. and Satz, H.",
    title = "{Lower Dimensional Gravity}",
    reportNumber = "MIT-CTP-1203",
    doi = "10.1016/0550-3213(85)90448-1",
    journal = "Nucl. Phys. B",
    volume = "252",
    pages = "343--356",
    year = "1985"
}

@article{Teitelboim1983,
    author = "Teitelboim, C.",
    title = "{Gravitation and Hamiltonian Structure in Two Space-Time Dimensions}",
    doi = "10.1016/0370-2693(83)90012-6",
    journal = "Phys. Lett. B",
    volume = "126",
    pages = "41--45",
    year = "1983"
}

@article{MaldacenaStanfordYang2016,
    author = "Maldacena, Juan and Stanford, Douglas and Yang, Zhenbin",
    title = "{Conformal symmetry and its breaking in two dimensional Nearly Anti-de-Sitter space}",
    eprint = "1606.01857",
    archivePrefix = "arXiv",
    primaryClass = "hep-th",
    doi = "10.1093/ptep/ptw124",
    journal = "PTEP",
    volume = "2016",
    number = "12",
    pages = "12C104",
    year = "2016"
}

@article{SaadShenkerStanford2019,
    author = "Saad, Phil and Shenker, Stephen H. and Stanford, Douglas",
    title = "{JT gravity as a matrix integral}",
    eprint = "1903.11115",
    archivePrefix = "arXiv",
    primaryClass = "hep-th",
    month = "3",
    year = "2019"
}

@article{StanfordWitten2017,
    author = "Stanford, Douglas and Witten, Edward",
    title = "{Fermionic Localization of the Schwarzian Theory}",
    eprint = "1703.04612",
    archivePrefix = "arXiv",
    primaryClass = "hep-th",
    doi = "10.1007/JHEP10(2017)008",
    journal = "JHEP",
    volume = "10",
    pages = "008",
    year = "2017"
}

@article{Marnelius:1982pf,
    author = "Marnelius, Robert",
    title = "{Polyakov's Spinning String From a Canonical Point of View}",
    reportNumber = "GOTEBORG-82-34",
    doi = "10.1016/0550-3213(83)90586-2",
    journal = "Nucl. Phys. B",
    volume = "221",
    pages = "409",
    year = "1983"
}

@book{Glimm:1987ylb,
    author = "Glimm, James and Jaffe, Arthur",
    title = "{Quantum Physics: A Functional Integral Point of View}",
    doi = "10.1007/978-1-4612-4728-9",
    isbn = "978-0-387-96477-5, 978-1-4612-4728-9",
    publisher = "Springer",
    year = "1987"
}

@article{Haag:1963dh,
    author = "Haag, Rudolf and Kastler, Daniel",
    title = "{An Algebraic approach to quantum field theory}",
    doi = "10.1063/1.1704187",
    journal = "J. Math. Phys.",
    volume = "5",
    pages = "848--861",
    year = "1964"
    }

@article{Buchholz:1981fj,
    author = "Buchholz, Detlev and Fredenhagen, Klaus",
    title = "{Locality and the Structure of Particle States}",
    reportNumber = "FREIBURG-THEP 81/6",
    doi = "10.1007/BF01208370",
    journal = "Commun. Math. Phys.",
    volume = "84",
    pages = "1",
    year = "1982"
}

@article{Gamboa:2025qjr,
    author = "Gamboa, J. and Mendez, F.",
    title = "{QED-IR as topological quantum theory of dressed states}",
    eprint = "2507.11668",
    archivePrefix = "arXiv",
    primaryClass = "hep-ph",
    doi = "10.1007/JHEP11(2025)040",
    journal = "JHEP",
    volume = "11",
    pages = "040",
    year = "2025"
}

@article{Gamboa:2025dry,
    author = "Gamboa, J.",
    title = "{Topology and the infrared structure of quantum electrodynamics}",
    eprint = "2505.13247",
    archivePrefix = "arXiv",
    primaryClass = "hep-th",
    doi = "10.1007/JHEP07(2025)184",
    journal = "JHEP",
    volume = "07",
    pages = "184",
    year = "2025"
}

@article{Gamboa:2025fcn,
    author = "Gamboa, Jorge",
    title = "{Entanglement and effective field theories}",
    eprint = "2502.11819",
    archivePrefix = "arXiv",
    primaryClass = "hep-th",
    doi = "10.1016/j.physletb.2025.139723",
    journal = "Phys. Lett. B",
    volume = "868",
    pages = "139723",
    year = "2025"
}

@article{HaagKastler,
  author = {R. Haag and D. Kastler},
  title = {An Algebraic Approach to Quantum Field Theory},
  journal = {J. Math. Phys.},
  volume = {5},
  pages = {848},
  year = {1964}
}

@article{Doplicher:1971wk,
    author = "Doplicher, Sergio and Haag, Rudolf and Roberts, John E.",
    title = "{Local observables and particle statistics. 1}",
    doi = "10.1007/BF01877742",
    journal = "Commun. Math. Phys.",
    volume = "23",
    pages = "199--230",
    year = "1971"
}

@article{Doplicher:1973at,
    author = "Doplicher, Sergio and Haag, Rudolf and Roberts, John E.",
    title = "{Local observables and particle statistics. 2}",
    doi = "10.1007/BF01646454",
    journal = "Commun. Math. Phys.",
    volume = "35",
    pages = "49--85",
    year = "1974"
}

\end{document}